\documentclass[a4paper]{article}
\usepackage{amsmath}
\usepackage{amsfonts}
\usepackage{graphicx}
\usepackage{geometry}
\usepackage{pifont}

\begin{document}

\title{A Predation Behavior Model Based on Game Theory}

 \author{Shi Chen \thanks{School of Biology Science,Nanjing University,Nanjing 210093,China Email:njuchenshi@gmail.com}
  ,Sheng Bao \thanks{Dept. of Information
Engineering,Nanjing Univ. of P.\&T.,Nanjing 210046, China, Email:forrestbao@yahoo.com.cn}
   ,Ling Yan\thanks{Dept. of Management Engineering, Nanjing Univ. of P.\&T.,Nanjing 210046, China}
   ,Cheng Huang\thanks{School of Biology Science, Nanjing University, Nanjing 210046,
 China}}

\date{}
 \maketitle

\begin{abstract}
This article adopts game theory to build a model for explaining the predation behavior of animals.We assume that both the prey and the preydator have two stratigies in this game,the active one and the passive one.By calculating the outcome and the income of energy in different stratigies, we find the solution to analyze the different evolution path of both the prey and the predator.A simulation result approximately represents the correctness of our model.
\end{abstract}

\section{Introduction}
The difference between plants
and animals is that animals have the ability to move
independently while plants do not. However, in the process of evolution, many animals
lose this ability and even become motionless. For instance, many
insects who own mimicry and protective color are stolid. On the other hand, some other animals, gain the extreme motive ability, as the felines show.
What we care about is which path is better for survival.Some other papers presented the basis on which we build our model. \cite{predator}\cite{fitness}

Game theory is usually used to solve economic and social problems. In this article,we use it  investigate the behavior of animal predation.

We assume the relationship between the predator
and the prey accords with the requirement of zero-sum game.
Concretely, what the predator gets from its prey is exactly what
the prey loses. Both the predator and the prey have two
strategies,the active one and passive one.We will discuss later
what these strategies are. We can infer the results of these four
combinations should be different. What we really care about is
that how the prey and the predator play this game to satisfy
both of them? Is there any equilibrium in the game? So we build up
a model to solve these problems.

\section{Preliminaries}

\subsection{Term Explaination}

\begin{enumerate}
\item Player:A player is an agent who makes decisions in a
game.\cite{Game}Here it means either the predator or the prey.

\item Strategy combinations: The predator has two ways to obtain food,to run fast in order to catch running prey or to stay quietly so that food will come automatically.The prey also has two ways to avoid being caught,running faster than the predator or hide themselves from the predator's sense.The motive one is considered as the active one while its opponent is considered as the passive one.So the strategy combination has four possibilities.

\item Payoff: In this article,the payoff is the algebra sum of energy outcome and income.Thus, the net income of the predator.
\end{enumerate}

\subsection{Assumptions}

\begin{enumerate}
\item We do not care the relationship between animals and plants,
we only focus on the predation relationship between animals.

\item This game is a zero-sum game which means the income
of the predator would be the outcome of the prey.
 \item Each play in this game is rational.
\end{enumerate}

\section{Building the model}

\subsection{Payoff of Predator}

According to the assumption of Shoener\cite{Lucas}, the income and outcome of
energy of predator consist of three parts:

\begin{enumerate}
\item Basic daily energy$E_{r}=k_{1}B^{3/2}$

\item The contained energy per unit of food$E_{i}=k_{2}i^{2}$.Since the obtained food
is$rk_{4}n$,the total incoming energy from food should be$\sum E_{i}=rk_{4}nk_{2}i^{2}$

\item The energy cost on running$E_{p}=k_{3} B^{3/2} r/v $. Note: Passive
predator does not have this energy output.

\end{enumerate}

The variations above have biological meanings as follows:

r:The distance that a predator covers

k:Probability of finding a prey

n:Probability of killing a prey

v:Velocity of a predator

i:Scale of the prey (be porprotional to body weight)

B:Scale of the predator (be porprotional to body weight)

$k_{1}$,$k_{2}$,$k_{3}$are parameters depend on circumstances.

Thus only when $\sum E_{i} - (E_r + E_p ) \ge 0$ the predator are able to
guarantees its energy income great than outcome.In other words, when the
payoff is positive, the killing is effective. Thus
$$ rnk_{2}k_{4}i^{2}-(k_{1} b^{3/2}+k_{3} B^{3/2}r/v) \ge 0$$

\subsection{Payoff matrix}

Considering two strategies of both the predator(Player I)and
the prey(Player II), we obtain the payoff table
\begin{table}[htb]
\begin{center}
\begin{tabular}
{|c|c|c|}
\hline & Escape(Active)& Lurk(Passive) \\
\hline Pursuit(Active)& e$_{1}$& e$_{2}$ \\
\hline Ambush(Passive)& e$_{3}$& e$_{4}$ \\
\hline
\end{tabular}
\end{center}
\label{tab1} \caption{Payoff Table}
\end{table}

Then we obtain the payoff matrix
 $$\left(
\begin{array}{cc}
e_{1} & e_{2} \\
e_{3} & e_{4}%
\end{array}%
\right) $$

where e$_{i}$(i=1,2,3,4) is the payoff of the predator in different strategies.

Consider the pursuit-escape strategy combination which is the simplest condition.The payoff of predator is
\begin{equation}
e_{1}=rnk_{2}ki^{2}-(k_{1}B^{3/2}+k_{3}B^{3/2}r/v)
\label{e1}
\end{equation}

Then,let us discuss the ambush-escape strategy combination.The predator has no energy consumption in pursuit since the prey comes into its mouth instead.Its payoff is
$$
e_{3}=r\textsf{'}nk_{2}ki_{2}-k_{1}B^{3/2}
$$

 where r\textsf{'}
is the distance that the prey covers. r\textsf{'} could be considered as the
predator's pursuit distance.Thus $r=r\textsf{'}$So the formula can be rewriten as
\begin{equation}
e_{3}=rnk_{2}ki_{2}-k_{1}B^{3/2}
\label{e2}
\end{equation}

To pursuit-lurk strategy combination,because the prey uses hiding strategy, we assume the
probability of discovering the prey decreases to $k-\alpha$.
But since the prey gives up escaping, we assume the killing probability then rises to $ n+\beta$. The payoff in this condition is
\begin{equation}
e_{2} =r(n + \beta )k_{2} (k - \alpha )i^2 - (k_{1}B^{3/2}+k_{3}B^{3/2}r/v)
\label{e3}
\end{equation}


In the case of the ambush-lurk strategy combination, both the prey and the predator stay still. The predator would starve to death because no food is available. Thus
\begin{equation}
e_{4}=0
\label{e4}
\end{equation}

\subsection{Biological meaning of the payoff matrix}
The meaning of mixed strategies is: for a specified
predator, we calculate many predation
process and see in how many times them use active strategy and how many times they use the
passive one in order to determine the evolutionary path of the predator.

\subsection{Solution of mixed strategy}
According to the game theory,the solution of mixed strategy is the solution of the following equation

\begin{equation}
xe_{1}+(1-x)e_{3}=xe_{2}+(1-x)e_{4}
\label{solution}
\end{equation}
where x is the probability that the predator use active strategy.

Substituting eq.\ref{e1},eq.\ref{e2},eq.\ref{e3}and eq.\ref{e4} into eq.\ref{solution},we obtain
$$x=\frac{rnkk_{2}i^{2}-k_{1}B^{3/2}}{r(n+\beta )k_{2}(k-\alpha
)i^{2}-k_{1}B^{3/2}}$$

Here we find an interesting phenomenon.The parameter $k_{3}$ disappears. Thus,when $r,n,k,k_{1}$ and $k_{2}$ are fixed,x is just a function of i and B.

Meanwhile, the form of numerator and denominator
is extremely similar which suggests the range of x is not quite large. So the proper$\alpha $and$\beta $ are of most
importance.

\section{An approximate test}
\begin{figure}[!htb]
\centering
\includegraphics[scale=0.6]{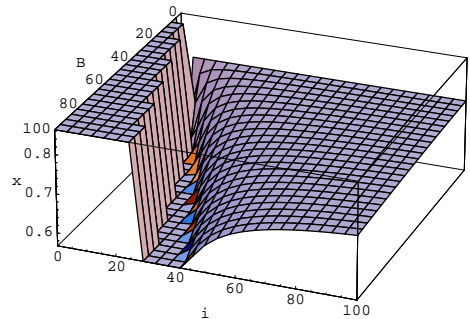}
\caption[example]{Test 1,$r=2000$,$n=0.3$,$k=0.8$,$v=10$,$k_{1}=500$,$k_{2}=1$,$\alpha=0.2$,$\beta=0.1$}
\label{iBx}
\end{figure}

Since lack of data, we can not compare our theory to real
condition. We perform a test based on given values. The
relationship between i,B and x is illustrated in Fig.\ref{iBx}

%


\section{Discussion}
The result from the model is similar with the nature. We cannot see a
spider as large as a leopard hiding in the woods and weaving its
web.A such large killer must walk around or chase its victims.
Conversely, the small killer more or less chooses ambush strategy.

After reading the publication of Kleibler which revealed the
velocity of mass reduction of a hungry animal is directly
proportional to its metabolism rate.And the Chossat law said that
an animal would die if it uses more than half body mass without
supply in time.We can see clearly those small animals cannot
suffer hunger as the larger ones. Consequently, the smaller
hunters cannot tolerate the result of a series of failure in
predation, which means death.

Yet to the smaller predators, passive strategy is a good choice.
On one hand, it guarantees the maximum energy income(while at the
same time at a high stake).On the other hand, the energy outcome
equals as low as zero. Combining these two advantages we could not
say it is an undesirable choice that so many predators hold, which
then annihilate their ability of fast moving.

Nonetheless, to larger predators, it is more difficult for
them to hide from the eyesight of preies. Thus no huge payoff
offered by passive hiding strategy exists any more. A careful
scrutiny of the payoff matrix represents that to large predators, their $e_{1}$
almost equals $e_{3}$, making passive strategy useless.


Finally, we must point out although this article is about game
theory and its biological application, we cannot forget the basic
truth that as of today, no evidence could indicate an animal can actively
select different strategies.The so-called mixed strategies
are the result of long time evolution and nature selection. We
hold the belief that there must be a mathematical optized solution
behind any biological behavior.

\section{Acknowledgement}
We appreciate professor Jianxiu Chen and associate professor
Zhengxiang He,both of who come from School of Biology Science, Nanjing
University, for their instructive and helpful discussions.

\bibliographystyle{IEEETran}
\bibliography{prey}

\end{document}